# Communication-Efficient Edge AI Inference Over Wireless Networks

YANG Kai, ZHOU Yong, YANG Zhanpeng, SHI Yuanming

(School of Information Science and Technology, ShanghaiTech University, Shanghai 201210, China)

**Abstract**
**Given the fast growth of intelligent devices, it is expected that a large number of high-stake artificial intelligence (AI) applications, e.g., drones, autonomous cars, tactile robots, will be deployed at the edge of wireless networks in the near future. As such, the intelligent communication networks will be designed to leverage advanced wireless techniques and edge computing technologies to support AI-enabled applications at various end devices with limited communication, computation, hardware and energy resources. In this article, we shall present the principles of efficient deployment of model inference at network edge to provide low-latency and energy-efficient AI services. This includes the wireless distributed computing framework for low-latency device distributed model inference as well as the wireless cooperative transmission strategy for energy-efficient edge cooperative model inference. The communication efficiency of edge inference systems is further improved by building up a smart radio propagation environment via intelligent reflecting surface.**

**Keywords**
**Communication efficiency; cooperative transmission; distributed computing; edge AI; edge inference.**

## 1 Introduction

The past few decades have witnessed a rapidly growing interest in the area of artificial intelligence (AI), which has contributed to the astonishing breakthroughs in image recognition, speech processing, etc. With the advancement of mobile edge computing [1], it becomes increasingly attractive to push the AI engine from the cloud center to the network edge. Such a transition makes AI proximal to the end devices and has the potential to mitigate the privacy and latency concerns. This novel area is termed as *edge AI*, including both edge training and edge inference, which is envisioned to revolutionize the future mobile networks and enable the paradigm shift from "connected things" to "connected intelligence" [2]. Edge AI can provide various AI services such as Internet of vehicles (IoVs), unmanned aerial vehicles (UAVs), and tactile robots, as illustrated in Figure 1. By deploying AI models and performing inference tasks at network edge, edge inference is the main focus of this article and faces the following three major challenges. First, the large size of AI models makes it difficult to be deployed at the network edge. Second, the inference latency is severely bottlenecked by the limited computation and communication resources at the network edge. Third, edge devices are usually battery-powered with limited energy budget and computing power.

It is generally impractical to deploy the entire AI models on a single resource-constrained end device. Fortunately, a recently proposed edge inference architecture, termed as *on-device distributed AI inference*, is capable of pooling the computing resources on a large number of distributed devices to perform inference tasks requested by each end device [3]. For popular distributed computing structures such as MapReduce [4], the dataset (i.e., AI model for inference) is split and deployed on end devices during the *dataset placement* phase. Each end device computes the intermediate values of all tasks locally with the *map* functions. After exchanging the intermediate values, each device obtains all map function values for its inference task, and performs the *reduce* function to yield the desired inference result. However, the communication efficiency of the intermediate value exchange is the main performance bottleneck of distributed edge inference systems [5]. To this end, we shall propose a communication-efficient data shuffling strategy for on-device distributed AI inference based on cooperative transmission and interference alignment.

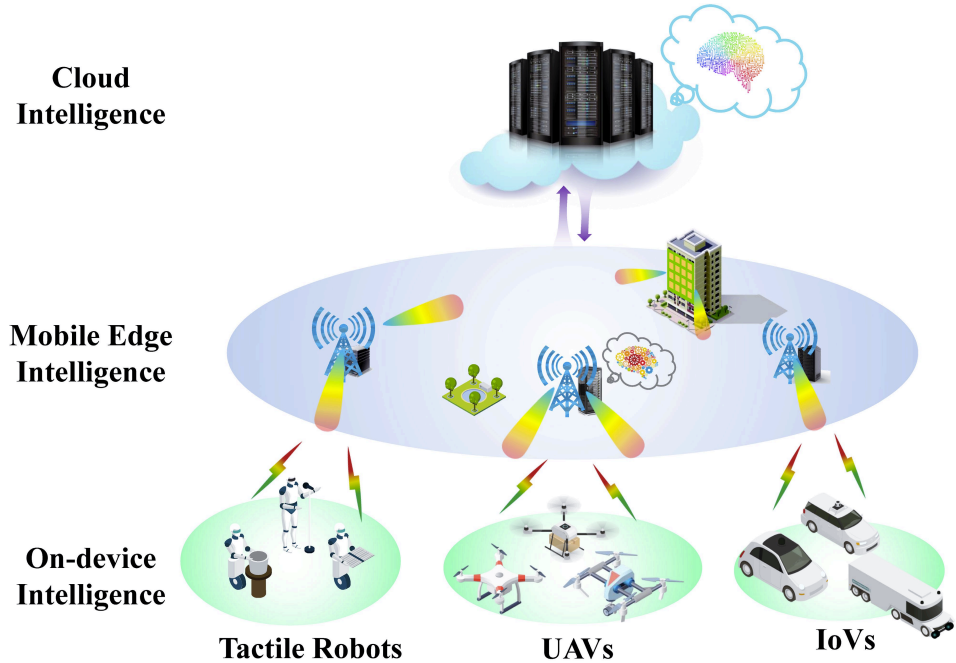

Figure 1: Illustration of edge AI.

For computation-intensive inference tasks, it is beneficial to deploy the AI models at the edge servers (e.g., access points (APs)), followed by uploading the input dataset to the proximal edge servers. This helps to perform the inference tasks and return the inference results to the end devices through downlink transmission. On the other hand, cooperative transmission [6] is a well-known approach that can mitigate co-channel interference as well as improve the reliability and energy efficiency of downlink transmission. These facts motivate us to propose the *in-edge cooperative AI inference* architecture by performing each task at multiple edge servers and enabling cooperative transmission to improve the quality-of-service (QoS) and reliability for inference results delivery. However, performing each inference task by multiple edge servers leads to a higher computation power consumption. We thus propose a joint task allocation and downlink coordinated beamforming approach to achieve energy-efficient in-edge cooperative AI inference through minimizing the total power consumption consisting of both transmit and computation power consumptions under the target QoS constraints.

Although our joint computation and communication designs can greatly improve the communication efficiency for on-device distributed AI inference and in-edge cooperative AI inference, the achievable low-latency and energy efficiency are still fundamentally limited by the radio propagation environment. We thus resort to an emerging technology, i.e., intelligent reflecting surface (IRS) [7], to actively control the wireless propagation environment. In particular, we propose to utilize the IRS for further enhancing the communication efficiency of edge AI inference systems, thereby providing low-latency and energy-efficient AI services. By dynamically adjusting the phase shifts of the IRS, our proposed strategy improves the feasibility of the interference alignment conditions for the data shuffling of on-device distributed AI inference systems, as well as reduces the energy consumption of in-edge cooperative AI inference systems.

## 2 Overview of Edge AI Inference

In this section, we present the architectures, key performance metrics, and promising applications of edge AI inference.

### 2.1 Architecture

In the conventional cloud-based AI systems, a large amount of data collected/generated by the end devices are required to be delivered to the central cloud center for AI model training. Such cloud-based AI systems are generally limited by scarce spectrum resources and susceptible to data modification attacks. With the increase of the computing power and storage capability of edge servers (e.g., APs and base stations) and end devices (vehicles and robots), there is a trend of pushing AI engines from the cloud center to the network edge [8], [9]. Therefore, edge AI emerges as a promising research area that performs the training and inference tasks at the network edge. In this article, we go beyond that and focus on a much broader scope of edge AI to fully leverage the distributed computation and storage resources at the network edge across end devices, edge servers and cloud centers to provide low-latency and energy-efficient AI services, such as IoVs, UAVs, and tactile robots.

According to [8], the edge AI can generally be classified into six levels, including cloud-edge co-inference, in-edge co-inference, on-device inference, cloud-edge co-training, all in-edge, and all on-device. The training of AI models can be performed on end devices, in edge servers, or with the collaboration of the cloud center and edge nodes, which are out of the scope of this article. We in this article mainly focus on the model inference of edge AI, also known as *edge inference*. The major architectures of edge inference are listed as follows:

- **Device-based edge inference:** Deploying AI models directly on end devices can reduce the communication cost due to information exchange. However, this poses stringent requirements on the storage capability, computing power, energy budget of each end device. To this end, a promising structure of device-based edge inference is to enable cooperation among multiple devices via a distributed computing framework [3], i.e., on-device distributed AI inference.
- **Edge-based edge inference:** The end devices offload the dataset to the neighboring edge servers, which perform the inference tasks and return the inference results to the end users. This inference architecture has the potential to perform computation-intensive inference tasks. However, the limited channel bandwidth is the main performance-limiting factor of this edge inference architecture. To address this issue, it is promising to enable cooperation among multiple edge servers [10], [11] to facilitate in-edge cooperative AI inference.
- **Others:** In addition to the device-based and edge-based edge inference architectures, there are also other promising edge inference architectures. The *device-edge architecture* with model partition proposed in [12] and [13] can enhance the energy efficiency and reduce the latency of edge inference systems. Moreover, the inference tasks can also be accomplished by adopting the *edge-cloud collaborative architecture*, which is particularly suitable for end devices with highly constrained resources.

This paper emphasizes on two promising system architectures, i.e., *on-device distributed AI inference* and *in-edge cooperative AI inference*, which pool the computation and communication resources across multiple end devices and edge servers, respectively. In such distributed systems, the communication efficiency is a critical issue in determining the performance of edge inference systems. We thus focus on designing communication-efficient on-device distributed AI inference and in-edge cooperative AI inference strategies for computation-intensive inference tasks, thereby achieving low latency and high energy efficiency.

## 2.2 Key Performance Metrics
The communication efficiency of edge inference systems can be measured by the following metrics:
- **Latency:** In edge inference systems, latency is a crucial performance metric that measures how fast the inference results can be obtained, which in turn determines the timeliness of the inference results. The latency is generally composed of the computation and communication latency. Achieving low latency is challenging as it depends on various factors, including channel bandwidth, transmission strategy, and channel conditions.
- **Energy efficiency:** As performing inference tasks is generally energy consuming, the energy efficiency is a critical performance metric of edge inference systems. The energy consumption typically consists of both communication and computation energy consumptions, which depend on the type of the inference tasks and the size of the dataset.
- **Others:** There are also other indicators that can describe the performance of edge inference. For example, *privacy* is a major concern in edge inference systems for various high-stake AI applications such as IoVs and UAVs. For such applications, it is also critical to ensure that the inference results are received at the end devices with a high level of *reliability*.

## 2.3 Applications
Efficient edge inference is envisioned to be capable of supporting various low-latency AI services, including IoVs, UAVs, and tactile robots, as shown in Figure 1.
- **Internet of vehicles:** IoV is a network system that integrates networking and intelligence for promoting the efficiency of transportation and improving the quality of life [14], as well as emphasizes the interaction of humans, vehicles, and roadside units. Numerous AI models are necessary for IoV such as the advanced driver-assistance system (ADAS) for the detection of vehicles, pedestrians, lane lines, etc. It is generally impractical to deploy all AI models on the resource-constrained vehicles. As a result, to achieve low-latency and energy-efficient inference for a large number of AI models, it is critical to pool the distributed computation and storage resources of the vehicles and edge servers at the network edge.
- **Unmanned aerial vehicles:** There has been a fast-growing interest in UAVs [15] for the transportation of cargo, monitoring, relaying, etc. Although the UAVs are battery-powered with limited energy budget, they are deployed to accomplish a variety of intelligent computation tasks. As it is energy inefficient for UAVs to communicate with the remote cloud center, enabling cooperative inference on the devices or in the edge is a promising solution that can achieve low-latency and energy-efficient processing of inference tasks, as well as enhance the data privacy.
- **Tactile robots:** As remote representatives of human beings, smart robots are envisioned to be capable of achieving physical interaction by enabling haptic capabilities, leading to the new field of *tactile robot* [16]. The greatly improved capability of processing tactile sensation and the connectivity of a large number of robots make tactile robots a representative embodiment of the tactile Internet. Exploring the potential of edge inference for tactile robots is able to provide integrated intelligence for agriculture, manufacture, health care, traffic, etc.

# 3 Wireless Distributed Computing System for On-Device Distributed AI Inference

In this section, we shall present a communication-efficient data shuffling strategy in the wireless distributed computing system for on-device distributed AI inference.

## 3.1 MapReduce-based Distributed Computing System

MapReduce is a ubiquitous distributed computing framework that processes tasks with a large amount of data across multiple distributed devices [4]. For a computing task with the MapReduce-like structure, the target function is decomposed as the *reduce* function value of a number of *map* functions, which can be computed in a parallel manner. Hence, the MapReduce-based distributed computing system is capable of pooling the computation and storage resources of multiple devices to enable on-device distributed AI inference.

For a wireless distributed computing system consisting of multiple mobile devices, the inference result (e.g., a machine learning model) to be obtained by each device depends on the entire input dataset. Supposing that each computation task for inference fits the MapReduce computation structure, as shown in Figure 2, $K$ mobile devices cooperatively accomplish the inference tasks through the following four phases

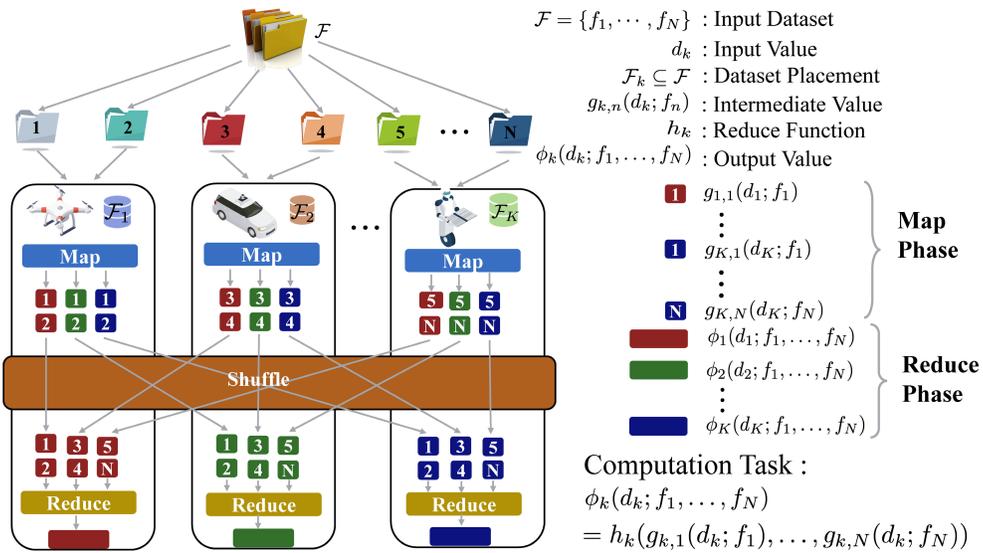

Figure 2: Illustration of computing model of MapReduce-based distributed computing framework.

- **Dataset placement:** In this phase, the entire dataset is partitioned into $N$ portions and each mobile device is allocated a subset of the entire dataset before inference.
- **Map:** With the allocated local data, each mobile device computes the map function values with respect to all the input data, which yields the intermediate values for itself and other devices.
- **Shuffle:** As each mobile device does not have enough information for inference, the intermediate values computed by each device shall be transmitted to the corresponding devices over radio channels in this phase.
- **Reduce:** Finally, based on the collected $N$ intermediate values, each mobile device calculates the reduce function to obtain the corresponding inference result.

With limited radio resources, the shuffling of intermediate values among multiple mobile devices leads to significant communication overhead and is the main performance-limiting factor for on-device distributed AI inference systems.

## 3.2 Communication-Efficient Data Shuffling

As data shuffling over radio channels is the major bottleneck of MapReduce-based distributed computing systems, it is necessary to propose a communication-efficient data shuffling strategy for a given dataset placement. Consider a wireless communication system consisting of multiple mobile devices and an AP, as illustrated in Figure 3. The basic idea for achieving low-latency data shuffling is to explore the opportunity of concurrent transmission, detailed as follows.
- **Uplink multiple access:** After computing the intermediate values with map functions, each mobile device transmits its precoded intermediate values to the AP over the multiple access channel.

- **Downlink broadcasting:** The AP broadcasts the received signal of uplink transmission to each device, which decodes its desired intermediate values.

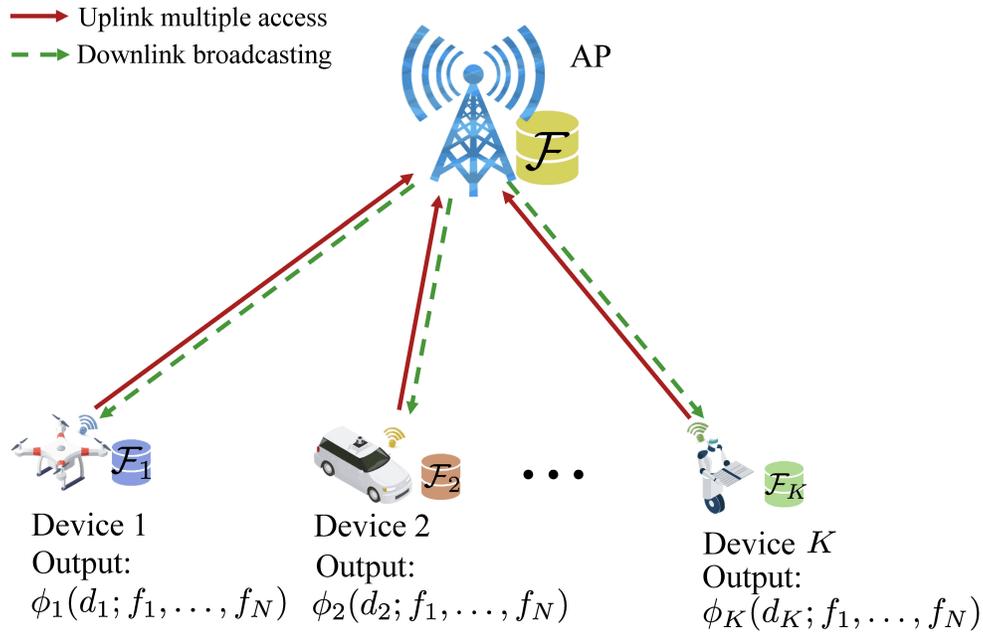

Figure 3: Illustration of communication model for data shuffling of on-device distributed AI inference systems.

The output of each computation task depends on not only the locally computed intermediate values at each device based on its own dataset but also intermediate values computed by other devices. By treating each intermediate value as an independent message, the data shuffling procedure is indeed a message delivery problem. The AP first receives a mixed signal from all mobile devices in the uplink, and then simply broadcasts the mixed signal to all mobile devices in the downlink. By studying the input-output relationship from all mobile devices to all mobile devices after the uplink and downlink transmissions, the proposed data shuffling strategy can be equivalently modeled as a data delivery problem over the $K$-user interference channel with side information available at both the transmitters and the receivers. Note that the AP behaves like a two-way relay [17] and simply transmits an amplified version of the received signal. The side information refers to the available intermediate values at each device. As a result, the goal becomes the transceiver design for maximizing the communication efficiency of data shuffling. It has been demonstrated that the linear coding schemes are effective for the transceiver design because of their optimality in terms of the degree-of-freedoms (DoFs) for interference alignment as well as low implementation complexity. Note that DoF is a first-order characterization of channel capacity, which is thus chosen as the performance metric for data shuffling. With interference alignment, the solutions meeting interference alignment conditions yield transceivers that are able to simultaneously preserve the desired signal and cancel the co-channel interference.

The problem of finding solutions to the interference alignment conditions with a maximum achievable DoF can be tackled by developing an efficient algorithm based on a low-rank optimization approach [3]. This is achieved by defining the product of the aggregated precoding matrix and the aggregated decoding matrix as a new matrix variable, based on the following two key observations
- The interference alignment conditions can be represented as affine constraints in terms of the newly defined matrix variable and the global channel state information.
- The rank of the matrix is inversely proportional to the achievable DoF.

Therefore, the maximum achievable DoF can be obtained via minimizing the matrix rank, subjecting to the affine constraints. For the nonconvex low-rank optimization problem, the traditional nuclear norm minimization approach yields unsatisfactory performance, which motivates us to propose a novel computationally efficient difference-of-convex-functions (DC) [17] algorithm to achieve considerable performance enhancement.

With limited radio resources, the scalability of the data shuffling strategy is also critical to the wireless distributed computing framework. We prefer a data shuffling strategy if the communication cost (which can be measured by achievable DoF) does not increase too much with more involved mobile devices. We present simulation results to demonstrate the effectiveness of the proposed algorithm for data shuffling. In simulations, we consider a single-antenna system, where the dataset is evenly split into 5 files and each device stores up to 2 files locally. With the uniform dataset placement strategy, each file is stored by $2K/5$ mobile devices. The achievable DoFs averaged over 100 channel realizations are illustrated in Figure 4. Interestingly, the achievable DoF of the proposed DC approach remains almost unchanged as the number of devices increases, while the nuclear norm relaxation approach suffers from a severe DoF deterioration. This demonstrates the scalability of the proposed DC approach. The main intuition is that the collaboration opportunities are increased as each file can be stored at more devices, although more intermediate values are requested with more involved devices.

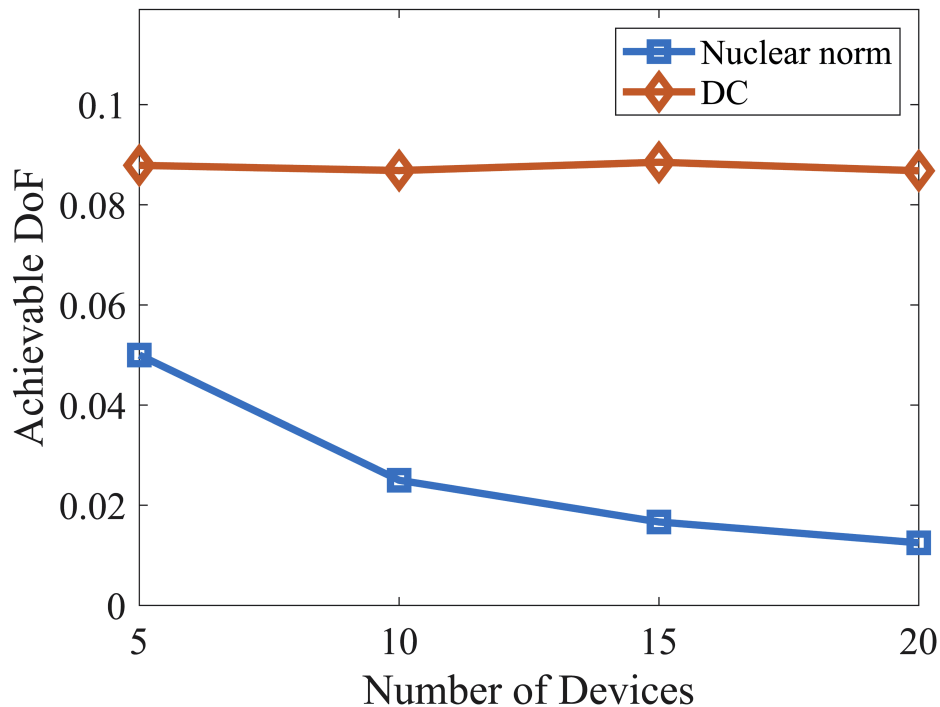

Figure 4: Achievable DoF of algorithms versus the number of devices for the data shuffling of on-device distributed AI inference systems.

## 4 Edge Processing System for In-Edge Cooperative AI Inference

In this section, we present a cooperative wireless transmission approach for energy-efficient edge processing of computational intensive inference tasks at edge servers.

Due to the strong capability of capturing data representation, machine learning techniques, in particular deep learning [19], have been widely used for achieving greatly improved performances in distilling intelligence from images, videos, texts, etc. However, the deep learning models are usually large and complex, and processing deep neural networks (DNNs) is a computation-intensive task. For resource-constrained mobile devices equipped with limited storage, computation power, and energy budget, such as drones and robots, a promising solution of performing computation-intensive inference tasks is to enable edge processing at the APs of mobile networks. With more powerful computing power than the resource-constrained mobile devices, APs have the potential of efficiently performing the inference tasks and transmitting the inference results to mobile users [21]. The design target is to enable cooperative transmission among multiple APs to provide higher quality-of-service (QoS) for reliably delivering the inference results, while minimizing the total power consumption consisting of the computation power of inference tasks and the transmission power at APs. The computation power of each task at an AP can be determined via estimating the energy consumption of processing DNNs [20] and computation time.

We consider a typical edge processing system consisting of $N$ APs served as edge processing nodes and $K$ mobile users, as demonstrated in Figure 5. Each mobile user has an inference task to be accomplished. The inference results can be obtained by uploading the input of each mobile user to the APs, processing a subset of inference tasks at each AP, and cooperatively transmitting the inference results the corresponding mobile user. The pre-trained models can be downloaded from cloud center and deployed at each AP in advance to facilitate edge inference. For example, the inference task can be the GauGan AI system by Nvidia, where the inputs are rough doodles and the outputs are photorealistic landscapes. Although the cooperative edge inference is able to deliver the reliable inference results to mobile users, the energy efficiency becomes critical as a huge amount of computations are required for processing DNNs at multiple APs.

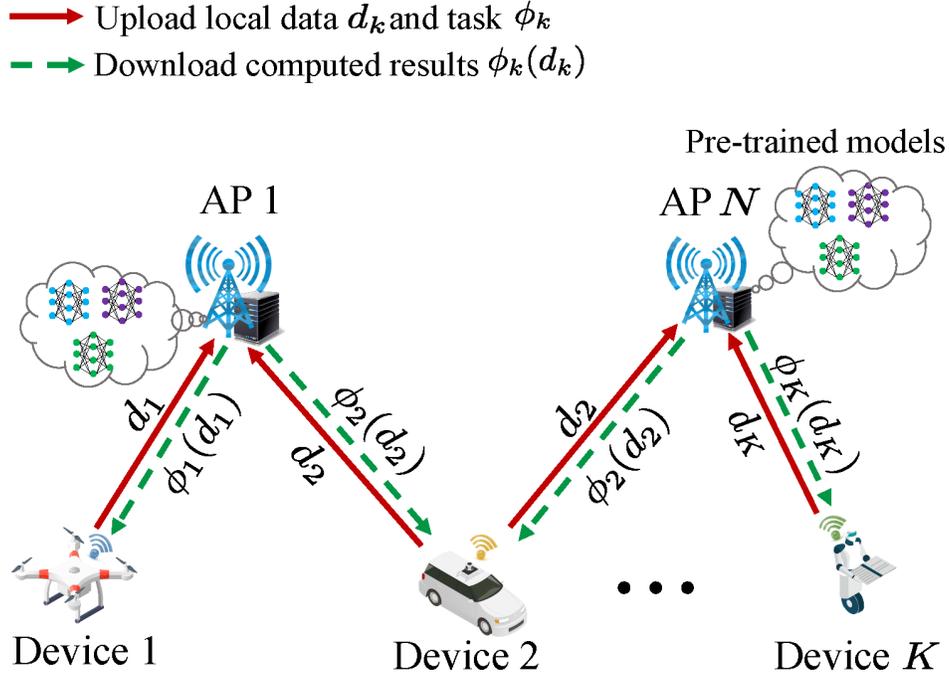

Figure 5: Illustration of in-edge cooperative AI inference systems.

There exists a tradeoff between communication and computation in an edge processing system by enabling cooperation among APs. In particular, if each AP performs more inference tasks, the inference results can be delivered with better QoS via cooperative downlink transmission. However, more computation power is consumed at the APs for processing the DNNs. To balance the tradeoff, we thus propose to minimize the total power consumption, consisting of computation power and communication power, under the target QoS constraints. This problem involves the joint design of the task allocation strategy across APs and the downlink beamforming vectors. Interestingly, if an inference task is not performed at one AP, the corresponding beamforming vector could be set as zero. This intrinsic connection between the task allocation strategy and the group sparsity structure of the downlink beamforming vectors allows us to reformulate the total power minimization problem under target QoS constraints as a group-sparse beamforming problem with QoS constraints. The group sparse structure can be induced with a well-recognized mixed $\ell_{1,2}$-norm, which results in a convex second-order cone program (SOCP) problem that can be efficiently solved. We leave simulation results in Figure 6 in Section 5.3 to evaluate the total power consumption of the proposed approach as well as the intelligent reflecting surface empowered in-edge cooperative AI inference.

## 5 IRS for Enhancing the Communication Efficiency of Edge Inference Systems

In this section, we introduce the novel intelligence reflecting surface (IRS) [8] technique for improving the signal propagation conditions of wireless environment, which is able to further enhance the communication efficiency for on-device distributed AI inference and in-edge cooperative AI inference.

**5.1 Principles of IRS**
An IRS is a low-cost two-dimensional surface of electromagnetic (EM) materials and composed of structured passive scattering elements [18]. The structural parameters determine how the incident radio waves are transformed at the IRS. The specially designed scattering elements introduce a shift of the resonance frequency and a change of boundary conditions, resulting in phase changes of both the reflected and diffracted radio waves. The scattering elements on IRS are reconfigurable by imposing external stimuli to alter their physical parameters, which can be exploited to fully control the phase shift of each element at the IRS.

Although the communication efficiency of data shuffling for on-device distributed AI inference, as well as wireless cooperative transmission for in-edge AI inference can be greatly improved by our novel communication strategy and algorithm design, it is still fundamentally limited by the wireless propagation environments. To this end, we resort to IRS that is capable of building a smart radio environment to address this issue.

**5.2 IRS-Empowered Data Shuffling for On-Device Distributed AI Inference**

IRS with real-time reconfigurability is capable of controlling the signal propagation environments, thereby improving the spectral efficiency and reducing the energy consumption of wireless networks. The controllable phase shifts to the incident signals makes IRS possible for further improving the achievable DoFs for data shuffling in Section 3. In particular, by actively reconfiguring the radio propagation environment, the feasibility of interference alignment conditions can be achieved. As a result, IRS is a promising technology for providing low-latency on-device distributed AI inference services for a wide range of applications. Note that we can still use the communication scheme and interference alignment technique provided in Section 3.2, and model the data shuffling problem as a side information aided message delivery problem in interference channel, while the channel coefficients could be adjusted by the phase shifts of IRS. The additional dimension provided by the phase shifts at RIS is able to further enhance the desired signals while nulling interference.

**5.3 IRS-Empowered In-Edge Cooperative AI Inference**

To further reduce the power consumption of in-edge cooperative inference in Section 4, it is promising to combat the unfavorable channel conditions by actively adjusting the phase shifts of IRS, rather than only adapting to the wireless propagation environments. By dynamically configuring the phase shifts of the IRS, a desired channel response can be achieved at the mobile devices, which in turn improves the signal power. Therefore, under the same QoS requirements, IRS can be utilized to further reduce the total power consumption of the edge processing system.

However, it calls for the joint design of the task allocation strategy, the downlink beamformers and the phase shifts of IRS. Exploiting the group structure of beamformers yields a highly nonconvex group-sparse optimization problem with coupled optimization variables in the QoS constraints, i.e., the downlink beamforming vectors at the APs and the phase shifts at the IRS. An alternating optimization framework can be adopted to decouple the highly nonconvex QoS constraints, for which updating the downlink beamforming vector is exactly the same as that in Section 4.1. The update for phase shifts at the IRS can be transformed to a homogeneous quadratically constrained quadratic program (QCQP) problem with nonconvex unit modulus constraints. To tackle the nonconvex constraints, the problem is further reformulated as a rank-one constrained optimization problem by leveraging the matrix lifting technique. The resulting optimization problem can then be solved with a DC algorithm by minimizing the difference between trace norm and spectral norm of the matrix variable.

For illustration purpose, we consider an edge processing system with three 5-antenna APs and ten single-antenna mobile users that are uniformly located in the square area of $[0, 200]$m $\times$ $[0,200]$m. An IRS equipped with 25 reflecting elements is deployed at the center of the square area. In simulations, the power consumption of performing an inference task at the AP is 0.45W and the maximum transmit power of AP is 1W. Figure 6 shows the average total power consumption versus the target signal-to-interference-plus-noise ratio (SINR) for edge processing systems without and with an IRS, where a simple random phase shift strategy is adopted. Simulation demonstrate that the power consumption can be significantly reduced by leveraging the advantages of IRS. We then compare the proposed DC approach with the semidefinite relaxation (SDR) approach as well as the random phase shifts strategy in Figure 7. It demonstrates that the proposed approach is able to achieve the least total power consumption among others.

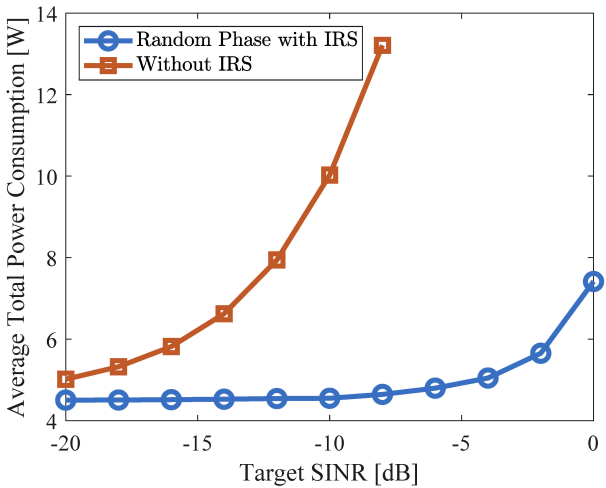
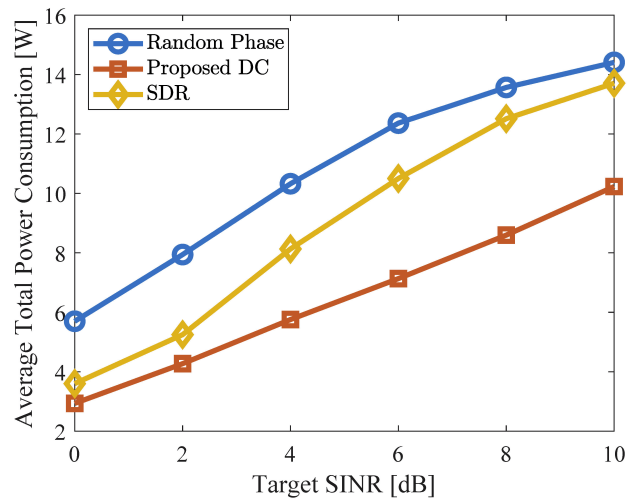

Figure 6: Average total power consumption comparison between edge processing systems with and without IRS.

Figure 7: Average total power consumption comparison for different algorithms in edge processing systems.

## 5 Conclusions

In this article, we presented the communication-efficient designs for edge inference. We identified two representative system architectures for edge inference, i.e., on-device distributed AI inference and in-edge cooperative AI inference. For on-device distributed AI inference, we proposed a low-latency data shuffling strategy, followed by developing a low-rank optimization method to maximize the achievable DoFs. We also proposed a group-sparse beamforming approach to minimize the total power consumption of in-edge cooperative AI inference. In addition, we explored the potential of deploying IRS to further enhance the communication efficiency by combating the detrimental effects of wireless fading channels. Our proposals are capable of achieving low-latency and high energy efficiency for edge AI inference.

## References

[1] Y. Mao, C. You, J. Zhang, K. Huang, and K. B. Letaief, "A survey on mobile edge computing: The communication perspective," *IEEE Commun. Surveys Tuts.*, vol. 19, pp. 2322–2358, Fourthquarter 2017.
[2] K. B. Letaief, W. Chen, Y. Shi, J. Zhang, and Y. A. Zhang, "The roadmap to 6G: AI empowered wireless networks," *IEEE Commun. Mag.*, vol. 57, pp. 84–90, Aug. 2019.
[3] K. Yang, Y. Shi, and Z. Ding, "Data shuffling in wireless distributed computing via low-rank optimization," *IEEE Trans. Signal Process.*, vol. 67, pp. 3087–3099, Jun. 2019.
[4] S. Li, M. A. Maddah-Ali, Q. Yu, and A. S. Avestimehr, "A fundamental tradeoff between computation and communication in distributed computing," *IEEE Trans. Inf. Theory*, vol. 64, pp. 109–128, Jan. 2018.
[5] S. Li, M. A. Maddah-Ali, and A. S. Avestimehr, "Coding for distributed fog computing," *IEEE Commun. Mag.*, vol. 55, pp. 34–40, Apr. 2017.
[6] D. Gesbert, S. Hanly, H. Huang, S. S. Shitz, O. Simeone, and W. Yu, "Multi-cell MIMO cooperative networks: A new look at interference," *IEEE J. Sel. Areas Commun.*, vol. 28, no. 9, pp. 1380–1408, 2010.
[7] X. Yuan, Y.-J. Zhang, Y. Shi, W. Yan, and H. Liu, "Reconfigurable-intelligent-surface empowered 6G wireless communications: Challenges and opportunities," *arXiv preprint arXiv:2001.00364*, 2020.
[8] Z. Zhou, X. Chen, E. Li, L. Zeng, K. Luo, and J. Zhang, "Edge intelligence: Paving the last mile of artificial intelligence with edge computing," *Proc. IEEE*, vol. 107, pp. 1738–1762, Aug. 2019.
[9] J. Park, S. Samarakoon, M. Bennis, and M. Debbah, "Wireless network intelligence at the edge," *Proc. IEEE*, vol. 107, pp. 2204–2239, Nov. 2019.
[10] K. Yang, Y. Shi, W. Yu, and Z. Ding, "Energy-efficient processing and robust wireless cooperative transmission for edge inference," *arXiv preprint arXiv:1907.12475*, 2019.
[11] S. Hua, Y. Zhou, K. Yang, and Y. Shi, "Reconfigurable intelligent surface for green edge inference," *arXiv preprint arXiv:1912.00820*, 2019.
[12] E. Li, L. Zeng, Z. Zhou, and X. Chen, "Edge AI: On-demand accelerating deep neural network inference via edge computing," *IEEE Trans. Wireless Commun.*, 2019.
[13] A. E. Eshratifar, M. S. Abrishami, and M. Pedram, "JointDNN: an efficient training and inference engine for intelligent mobile cloud computing services," *IEEE Trans. Mobile Comput.*, 2019.
[14] J. Zhang and K. B. Letaief, "Mobile edge intelligence and computing for the internet of vehicles," *Proc. IEEE*, 2019.
[15] Y. Zeng, Q. Wu, and R. Zhang, "Accessing from the sky: A tutorial on UAV communications for 5G and beyond," *Proc. IEEE*, vol. 107, pp. 2327–2375, Dec. 2019.
[16] S. Haddadin, L. Johannsmeier, and F. D. Ledezma, "Tactile robots as a central embodiment of the tactile internet," *Proc. IEEE*, vol. 107, pp. 471–487, Feb. 2019.
[17] K. Liu and M. Tao, "Generalized signal alignment: On the achievable DoF for multi-user MIMO two-way relay channels," *IEEE Trans. Inf. Theory*, vol. 61, no. 6, pp. 3365-3386, Jun. 2015.
[18] P. D. Tao and L. T. H. An, "Convex analysis approach to DC programming: Theory, algorithms and applications," *Acta Math. Vietnamica*, vol. 22, no. 1, pp. 289–355, 1997.
[19] S. Gong, X. Lu, D. T. Hoang, D. Niyato, L. Shu, D. I. Kim, and Y.-C. Liang, "Towards smart radio environment for wireless communications via intelligent reflecting surfaces: A comprehensive survey," *arXiv preprint arXiv:1912.07794*, 2019.
[20] Y. LeCun, Y. Bengio, and G. Hinton, "Deep learning," *nature*, vol. 521, no. 7553, p. 436, 2015.
[21] K. Li, M. Tao, and Z. Chen, "Exploiting computation replication for mobile edge computing: A fundamental computation-communication tradeoff study," to appear in *IEEE Trans. Wireless Commun.*, 2020. [Online]. Available: https://arxiv.org/abs/1903.10837.
[22] T.-J. Yang, Y.-H. Chen, and V. Sze, "Designing energy-efficient convolutional neural networks using energy-aware pruning," in Proc. IEEE Conf. Comput. Vision Pattern Recognition (CVPR), Jul. 2017, pp. 5687–5695.